\documentstyle[11pt,aaspp4,psfig]{article}

\received{\today}
\begin{document}
\title{Hard X-ray Lags in Cygnus X-1}
\author{D.J. Crary\altaffilmark{1}, M.H. Finger\altaffilmark{1},
        C. Kouveliotou\altaffilmark{1},
        F. van der Hooft\altaffilmark{2},\\ M. van der Klis\altaffilmark{2},
        W.H.G. Lewin\altaffilmark{3} and J. van Paradijs\altaffilmark{2,4}}

\altaffiltext{1}{Universities Space Research Association, Huntsville,
    AL 35806.}
\altaffiltext{2}{Astronomical Institute ``Anton Pannekoek'', University
    of Amsterdam \& Center for High-Energy Astrophysics, Kruislaan 403,
    NL-1098 SJ Amsterdam, The Netherlands.}
\altaffiltext{3}{Massachusetts Institute of Technology, 37-627 Cambridge,
    MA, 02139.}
\altaffiltext{4}{Department of Physics, University of Alabama in Huntsville,
    Huntsville, AL 35899.}

\begin{abstract}

We have used the Fourier cross spectra of Cyg X-1, as obtained with BATSE
during a period of almost 2000 days, to estimate the phase (or time) lags
between X-ray flux variations in the 20--50 keV and the 50--100 keV bands as a
function of Fourier frequency, $\nu$. We find that these lag spectra do not
depend on source state, as measured by the fractional rms variations of the
X-ray flux. For frequencies well below the Nyquist frequency the variations in
the 50--100 keV band lag those in the 20--50 keV band by a time interval $\tau
\propto \nu^{-0.8}$. Binning effects cause the  observed hard X-ray time lags
to decrease to zero at the Nyquist frequency. Some previous results on time
lags were affected by these binning effects. At photon energies above 10 keV
the time lags are approximately proportional to the logarithm of the ratio the
energies of the passbands used.

\end{abstract}


\centerline{To appear in Astrophysical Journal Letters}

\section{Introduction}

The delay of high-energy photons relative to low-energy photons in the X-ray
brightness variations of Cygnus~X-1 has been known for some time. The
significance of these lags is their possible use as a diagnostic of the source
of Compton upscattering of soft photons widely believed to be the origin of the
hard spectral tail seen in Cyg~X-1 and other black hole candidates (BHCs) 
(Miller\markcite{Miller} 1995; Nowak \& Vaughan\markcite{Nowak} 1996), as a
probe of the dynamics of the Comptonization region itself (Miyamoto et al.
\markcite{Miyamoto88} 1988), and as a second-order statistic supplementary to
the power spectrum in the determination of parameters in shot-noise models
(Miyamoto et al.\markcite{Miyamoto88} 1988).

The lags have been investigated using cross spectral analyses of the
time variations of the source in different energy channels (van der
Klis et al.\markcite{vdK95} 1995; see also 
Lewin, van Paradijs \& van der Klis\markcite{Lewin} 1988).
Miyamoto et al.\ (1988\markcite{Miyamoto88}, 1989\markcite{Miyamoto89}, 
1992\markcite{Miyamoto89}, 1993\markcite{Miyamoto93}) have applied this
technique to the BHCs Cyg~X-1, GX 339$-$4, GS 2023+338, and GS 1124$-$338
with Ginga data, using various combinations of
energy channels from 2--37 keV and a maximum sampling frequency of 125 Hz
(Nyquist frequency, $\nu_{\rm Nyq}$ = 62.5 Hz).
Their results showed that the lag spectra of these BHCs (i.e., lags as
a function of Fourier frequency) were very similar, severally for the 
low state and the very high state which were covered in their observations. 
The amplitude of the lag in general increases with photon energy, $E$,
except, perhaps, for $E \stackrel{<}{\sim} 5$ keV.
This consistent picture led Miyamoto et al.\markcite{Miyamoto92} (1992)
to refer to this lag properties as `canonical' for BHC sources.
The other type of X-ray binary for which the lag spectra have been
well studied are the Z sources. These contain accreting neutron stars;
they show different power spectral features, and have different lag
properties (van der Klis et al.\markcite{vdK87} 1987).

In general, the results of Miyamoto et al.\markcite{Miyamoto93} (1993) 
show a phase lag that 
increases as a function of frequency up to approximately $\sim$5 Hz, then
decreases toward $\nu_{\rm Nyq}$.  At the highest frequencies, the
uncertainties in the lag values become large enough to make the
determination of any trend uncertain.
The corresponding time lags decrease as the frequency
increases towards $\nu_{\rm Nyq}$
where they are consistent with zero. At least in Cyg~X-1,
the phase lags depend on the average energies $E_1$ and $E_2$ of the
energy bands used approximately as $\ln(E_2/E_1)$ 
(Miyamoto et al.\markcite{Miyamoto88} 1988).

We have used the approximately 2000 days of data from the Burst and Transient
Source Experiment (BATSE) on the {\it Compton Gamma Ray Observatory} to
investigate X-ray lags in variations from 
Cyg X-1, in a 
higher energy range than previously possible. Our data cover a large range 
in hard X-ray intensity and spectral slope (see Crary et al. 1996, and 
the X-ray light curves in Paciesas et al.\markcite{Paciesas97} 1997); 
our observations include time intervals in 
which an ultra-soft component is present; during these intervals Cyg X-1 
was likely in the high or intermediate state (see Zhang et
al.\markcite{Zhang97} 1997; 
Belloni et al.\markcite{Belloni96} 1996).
A description of
the data analysis is given in the next section. In sections 3 and 4
we compare our results to those of Miyamoto et al., and present
our conclusions.

\section{Observations}

We calculated lags between the 20--50 and 50--100 keV energy
bands of the $\Delta T=1.024$ s time resolution data obtained
with the large-area detectors (LADs). Fast Fourier transforms
were calculated in a way identical to that discussed in
Crary et al.\markcite{Crary96a}\ (1996a), 
with the addition of later observations.
These transforms were created for 524.288 second intervals (512 time bins)
in two energy channels, for all detectors with an angle
of less than $60^\circ$ to the direction of Cyg~X-1 (from one
to four detectors
view the source during a particular pointing period).  The average
number of uninterrupted 512 bin segments available with the source
unocculted by the Earth is approximately 40 per day.  The complex
cross amplitudes were created from the Fourier amplitudes
$a^i_j = \sum_k c^i_k \exp(i2\pi kj/n)$, where $n$ is the number
of time bins, $c^i_k$ is the detector count rate in bin $k =( 0\ldots n-1)$
and channel number $i = (1,2)$,
and $j = (-n/2\ldots n/2)$ corresponds to Fourier frequencies
$2\pi j/n\Delta T$.
The cross spectra were normalized to the total detector count rates
$N_i$ in channel $i$, by analogy with the method of 
Leahy et al.\markcite{Leahy} 1983),
and are given by
$C^{12}_j=<a^{2\ast}_j a^1_j/\sqrt{N_1 N_2}>$,
where the brackets refer to daily averaging.  
Errors on the averages of the real and imaginary parts of cross spectrum were
calculated from their respective sample variances. For further averages of
the cross spectrum these errors were propagated.
The phase lag as a function of frequency is obtained from the cross spectrum
as $\phi_j=\arctan{[{\rm Im}(C^{12}_j)/{\rm Re}(C^{12}_j)]}$. With the
above definitions, lags in the hard X-ray variations appear as
positive angles. 
For these data, deadtime induced cross-channel
effects (van der Klis et al.\markcite{vdK87} 1987) will affect only the
real part of the cross spectrum, reducing it by less then 5\%.

\section{Results}
\subsection{Calculation of Phase and Time Lags}
Because of the properties of the Fourier amplitudes  (van der Klis
\markcite{vdK89} 1989; Leahy et al.\markcite{Leahy} 1983)  and the noise in the
uncollimated BATSE detectors, the cross spectra for a large number of days 
must be averaged and converted to lag values to obtain sufficiently small 
errors. 
As shown by Crary et al.\markcite{Crary96a}\markcite{Crary96b}
(1996a,b) the slope of the X-ray spectrum of Cyg~X-1 and the amplitude of the
variability are strongly correlated. This indicates also that within the low
state the properties of Cyg~X-1 are determined by a simple parameter, most
likely the mass accretion rate.  We have therefore used the squared fractional
rms amplitude of the noise, integrated over the $0.03-0.488$ Hz frequency
range, as the relevant quantity with which to correlate the lag properties.
Figure 1b-d shows the data grouped into three parts, i.e.\ those for which the
daily averaged squared fractional rms values (between 0.03 and 0.488 Hz) are
greater than 0.03 and less that 0.05, between 0.05 and 0.07, and greater than
0.07, respectively. The figure shows that to within the uncertainties, there is
no obvious trend in the lag spectrum with source state.    
Figure 1a presents results obtained by averaging the cross spectra over the set
of data with fractional rms squared values greater than 0.03. 

\placefigure{fig1}

These data show that at the lowest frequencies, the frequency dependence of the
phase lag is consistent with zero phase lag; the phase lag rises to a peak
value of 0.04 radians near 0.2 Hz and decreases above this point to a value to
near zero radians at the $\nu_{\rm Nyq}$.  The resulting time lags, $\tau_j =
\phi_j/2\pi\nu_j$ (where $\nu_j$ is the frequency in Hz of the $j$th frequency
bin), are shown in Figure \ref{fig2} (filled circles).  These show a roughly
power-law decrease above $\sim\! 0.01$ Hz (exponent $\sim -0.8$) with a break
above 0.1 Hz, beyond which they decrease much more quickly.

\placefigure{fig2}

The shape of the phase lag spectrum looks superficially like
that found by
Miyamoto et al.\markcite{Miyamoto88} (1988) 
for Cyg~X-1 and by Miyamoto et al.\markcite{Miyamoto88} (1993)
for GS~1124-68 and GX~339-4. For the case of Cyg~X-1, however, the turn over in
their phase lag spectrum occurred at $\sim\! 30$ Hz, and these lags
decreased to near zero at their Nyquist frequency (62.5 Hz). The
similarity in the shape of these spectra over the sampling time dependent
frequency range suggests, by analogy with the finite-sampling effects
in power spectral analyses (van der Klis\markcite{vdK89} 1989), that at least
part of this shape (in particular the decrease toward the Nyquist
frequency) is determined by binning and aliasing effects on the cross
spectrum.

\subsection{Systematic Effects}
The effect of finite sampling on cross spectrum and lags can be calculated
in a way similar to the analogous problem in the determination of
power spectra (van der Klis\markcite{vdK89} 1989; see also 
Maejima et al.\markcite{Maejima} 1984).

The discrete form of the cross spectrum $C^{12}$
for the pair of binned, continuous random functions $x^j(t)$ with
zero mean (with $j$ equals 1 or 2) in terms of the cross spectrum
$\hat{C}^{12}$ of $x^1(t)$ and $x^2(t)$ is given by
\begin{displaymath}
C^{12}_l = \sum_{k=-\infty}^{\infty} \hat{C}^{12}
            (\frac{l}{n\Delta t}+\frac{k}{\Delta t})
            {\rm sinc}^2(\pi(\frac{l}{n}+k))
\end{displaymath}
where ${\rm sinc}\, x = \sin x/x$, $n$ is the number of
bins in a data segment, $\Delta t$ is the bin width, and
$l=-n/2\ldots n/2$. The sum over k represents the aliasing in the frequency
domain due to the uniform sampling in the time domain 
(van der Klis\markcite{vdK89} 1989), 
and the $sinc^2$ terms represent the effect of the averaging of the 
data over the finite time bin width. 

To investigate the effect of binning on the phase and time lags,
it is necessary to model the cross spectrum at
frequencies above $\nu_{\rm Nyq}$.
Here we assume a cross spectrum of the form
$\hat{C}^{12}(\nu)=F(\nu)e^{i2\pi\nu\tau(|\nu|)}$,
where $F(\nu)$, the modulus of the
cross spectrum, is independent of the detailed structure of
the time lag $\tau$ as a function of frequency, and the
difference between the real and imaginary parts of the cross
spectrum is determined by the complex exponential depending
on $\tau(|\nu|)$. This phenomenological model has the simple interpretation
that for $\tau$ independent of $\nu$ it represents a constant time lag.
The real and imaginary parts of the cross spectrum are plotted in
Figure \ref{fig3}.
The shape of the real part of the cross spectrum is suggestive
of the shape of a typical power spectrum, showing a `flat top' at
low frequency and a break at $\sim 0.1$ Hz to a power law form
at higher frequencies (Belloni \& Hasinger\markcite{Belloni} 1991; 
Crary et al.\markcite{Crary96b} 1996b).
To test the validity of the simple model for the cross spectrum, we have
modeled $F(\nu)$ with a Lorentzian-type function
$F(\nu)=a_1/(1+(|\nu|/a_2)^{a_3})$, and assume that the time lag has a
simple power-law form, $\tau=\tau_0|\nu|^{-\alpha}$
from an inspection of the shape of the time
lags in Figure \ref{fig2} at a point well away from $\nu_{\rm Nyq}$.
These five parameters were then determined by a simultaneous
fit of the real and imaginary parts of $C_l^{12}$ to the
cross spectral data.  The best fit parameters are found to be
$a_1=1.5$, $a_2=0.11$, $a_3=1.3$ for $F(\nu)$, and $\tau_0=0.0071$,
$\alpha= -0.78$ for $\tau(|\nu|)$.

\placefigure{fig3}

Using the binning correction to this cross spectrum model, the fit
to the data is shown as the dashed line in Figure \ref{fig3}.  Without
binning, the fit form is given by the dotted line in Figure \ref{fig3}.
Because of the small values of $\tau$ which are observed, the
fit values are insensitive to the exact functional from of $\tau(|\nu|)$,
but the binning correction is a large effect.  The effect of binning
on the time lags calculated from this model is shown in Figure \ref{fig2}
(dashed line). We find that the time lag, away from $\nu_{\rm Nyq}$ depends on
frequency, $\nu$, approximately as $\nu^{-0.8}$ up to $\nu \sim 0.2$Hz.
Above this frequency, the lags downturn, most of which is an artifact
due to the data binning.

It should be stressed that these results are not meant to provide a
definitive description of the cross spectra, but to illustrate the
important effect of binning on the
lag spectrum at frequencies above $\sim 0.5\nu_{\rm Nyq}$. In this model,
the calculation of the time lags is not sensitive to the the detailed
shape of $F(\nu)$, but it is clearly not adequately
described by a simple Lorentzian-like function.  Other potential systematic
effects on the cross spectrum, such as `windowing' in the
time domain (van der Klis\markcite{vdK89} 1989), have not been considered.
However, the implications of this simple
model are clear. At least part of the decrease in the time lags
with frequency observed by Miyamoto et al. (1992\markcite{Miyamoto92}, 
1993\markcite{Miyamoto88}) must be
attributed to binning effects.  To obtain a precise determination
of the phase or time lag as a function of frequency, these effects
must be considered, or, since this effect dominates near
the $\nu_{\rm Nyq}$ frequency, frequencies above
$\sim 0.5\,\nu_{\rm Nyq}$ should be ignored.

Since we find that the lags are independent of rms fractional X-ray flux
variations, we have combined our time lag results
with those obtained by Miyamoto et al.\markcite{Miyamoto93}  (1989) for Cyg
X-1. This allows us to study the photon energy dependence of the time lags over
a much increased energy range. Miyamoto et al.\markcite{Miyamoto88}  (1989)
found that the time lags between energy bands with average energies $E_2$ and
$E_1$ are proportional to $\epsilon = \ln (E_2/E_1)$. For power law photon
indices between $-1.5$ and $-2.5$ we find that for the BATSE energy bands
$\epsilon$ is in the range 0.802 and 0.833. To avoid the binning effects we
have taken the time lag at $\nu = 0.1$~Hz, for which we find $\tau = 0.055 \pm
0.005$~s. 
Then the proportionality constant 
between time lag and the logarithm of the energy ratio 
$\theta = \tau /\epsilon= 0.067 \pm 0.006$. 
At the same frequency we infer from Figure 3 
of Miyamoto et al.\markcite{Miyamoto88}  (1989)
lags of 0.045 s, 0.030 s, and 0.088 s 
between the 1.2--4.7 keV and
4.7--9.3 keV bands, between the 4.7--9.3 keV and 9.3--14.0 keV bands, and 
between the 4.7--9.3 keV and 15.8--24.4 keV bands, respectively. The
corresponding values of $\theta$ are 0.045, 0.060 and 0.080, 
respectively (estimated accuracy $\sim 10$ percent). 
This indicates that between the energy ranges below 10 keV and above 10 keV 
the ratio $\theta$ of the time lags to $\ln (E_2/E_1)$ increases; however, 
between $\sim 10$ and 100 keV $\theta$ does not change within the accuracy 
with which it has been determined.

\section{Conclusions}
Based on our analysis of nearly 2000 days of BATSE data we have
studied the time lags between the X-ray flux variations of Cyg X-1 in
the 20--50 keV and 50--100 keV bands as a function of Fourier frequency.
We find that the time lag spectra do not depend on source state as
measured by the rms fractional variations of the X-ray flux. Over the
range 0.01--0.2 Hz the hard lags $\tau$ are proportional to
$\nu^{-0.8}$. At frequencies above $0.5 \nu_{\rm Nyq}$ the lags are
artificially decreased (to zero at $\nu_{\rm Nyq}$) because of data
binning. At least part of the `canonical' time lag behavior of black
holes, as defined by Miyamoto et al., is due to this binning effect.
At photon energies above 10 keV the time lags are approximately proportional
to the logarithm of the ratio of the energies of the passbands used. 

\acknowledgments
This project was performed within NASA grant NAG5-2560 and supported
in part by the Netherlands Organization for Scientific Research (NWO)
under grant \mbox{PGS 78-277} and
by the Netherlands Foundation for Research in Astronomy (ASTRON)
under grant 781-76-017. Part of this work was completed while D.J.C held
a National Research Council-NASA Research Associateship.
F.v.d.H. acknowledges support by the Netherlands Foundation for Research in
Astronomy with financial aid from NWO under contract number
\mbox{782-376-011}.
W.H.G.L gratefully acknowledges support form the National Aeronautics
and Space Administration, and J.v.P acknowledges support from NASA
under contract \mbox{NAG5-2755} and \mbox{NAG5-3003}.

\newpage

\psfig{file=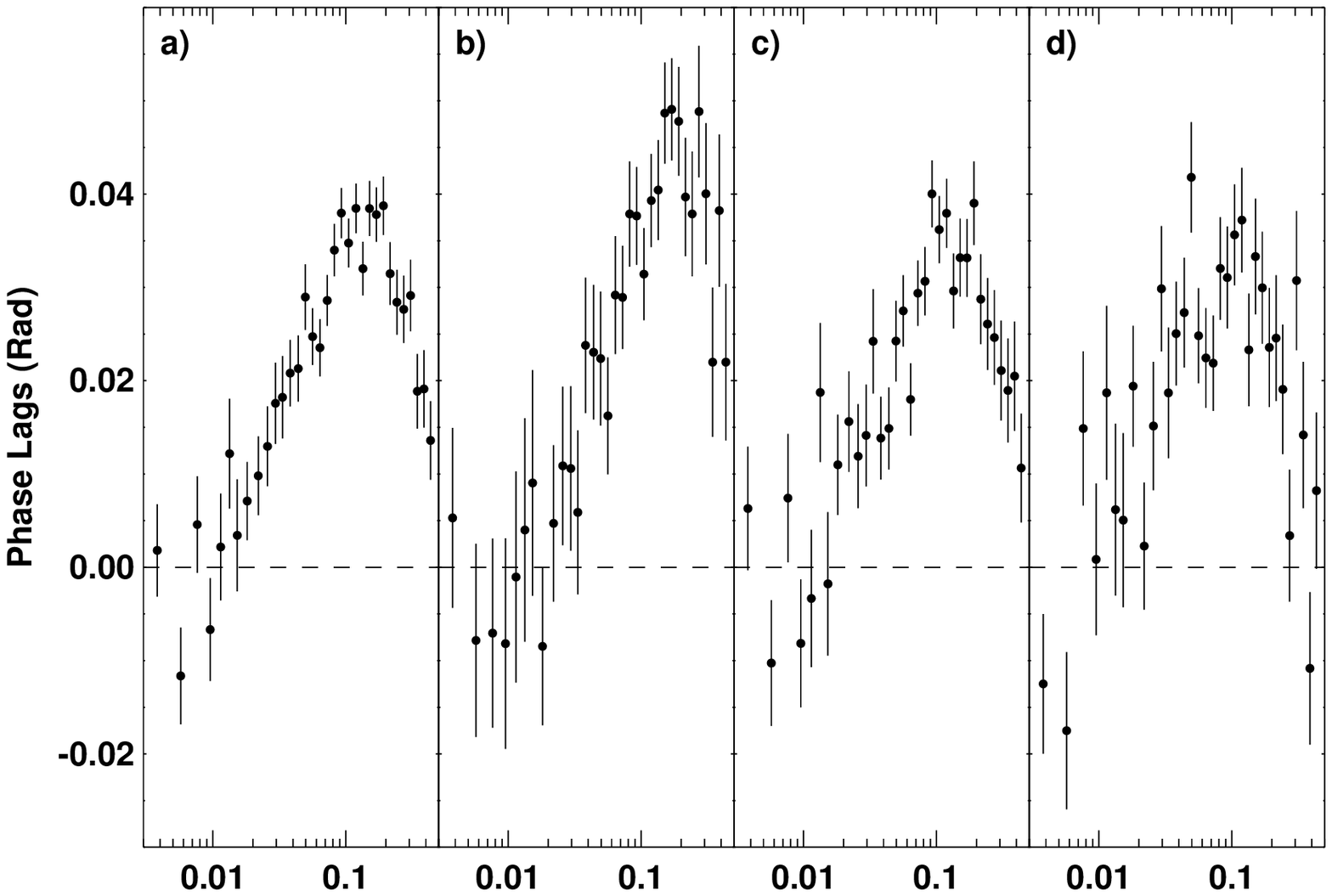,width=6.5in}
\figcaption[crary_fig1.ps]{Phase lags between 20-50 keV and 50-100 keV data for
various daily averaged squared rms levels, $s$.  a) All data with
$s > 0.03$,
b) $0.03 s < 0.05$, c) $0.05 < s < 0.07$, d) $0.07 < s$.\label{fig1}}

\psfig{file=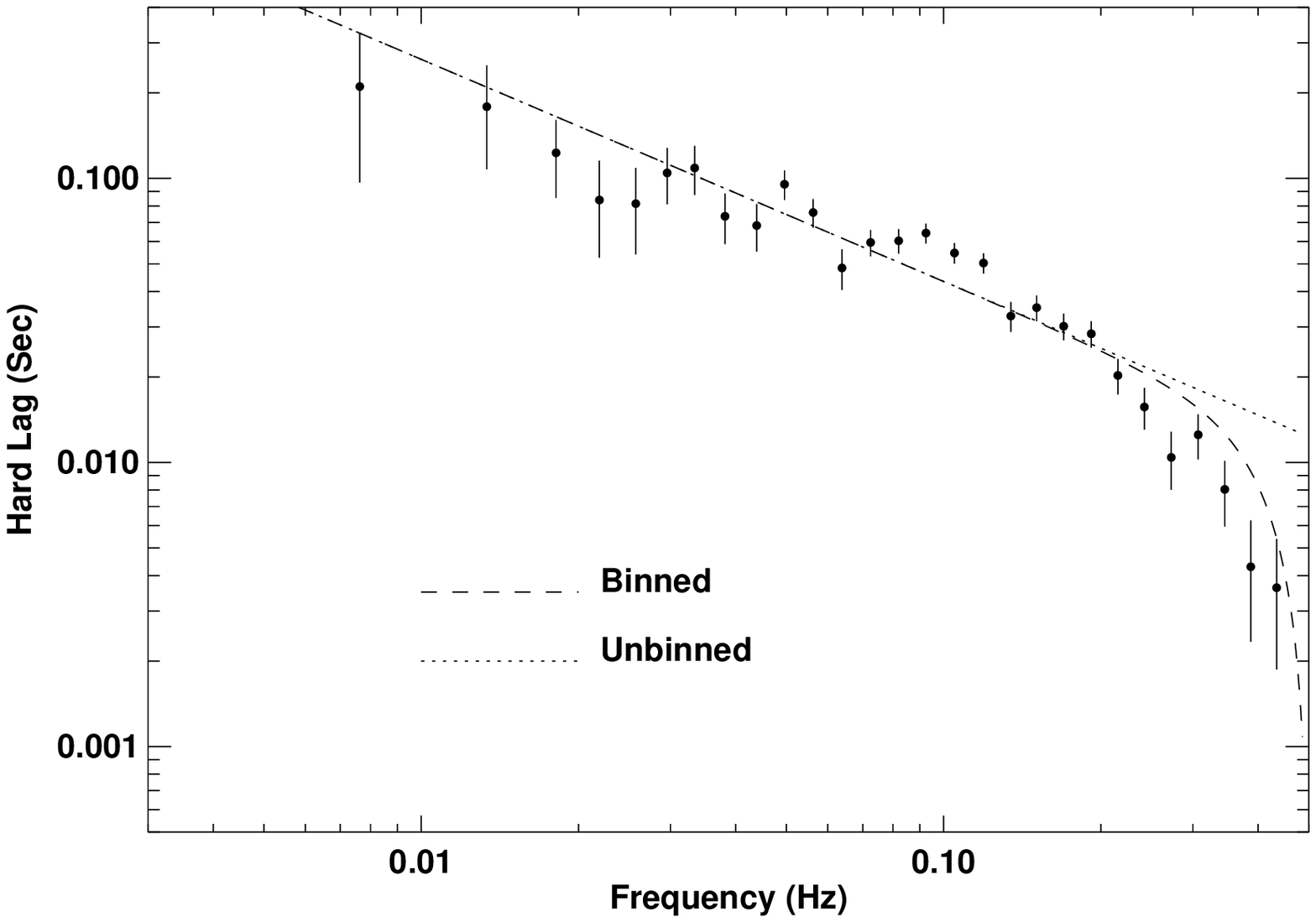,width=6.5in}
\figcaption[crary_fig2.ps]{Time lags as a function of frequency.
Solid circles denote values obtained by averaging over all data with
fractional squared rms values greater than 0.03.
The dashed line is the time lag obtained from a fit
to the lag model including binning effects (see text).  The
dotted line is a power law with parameters determined from
the model fit.\label{fig2}}

\psfig{file=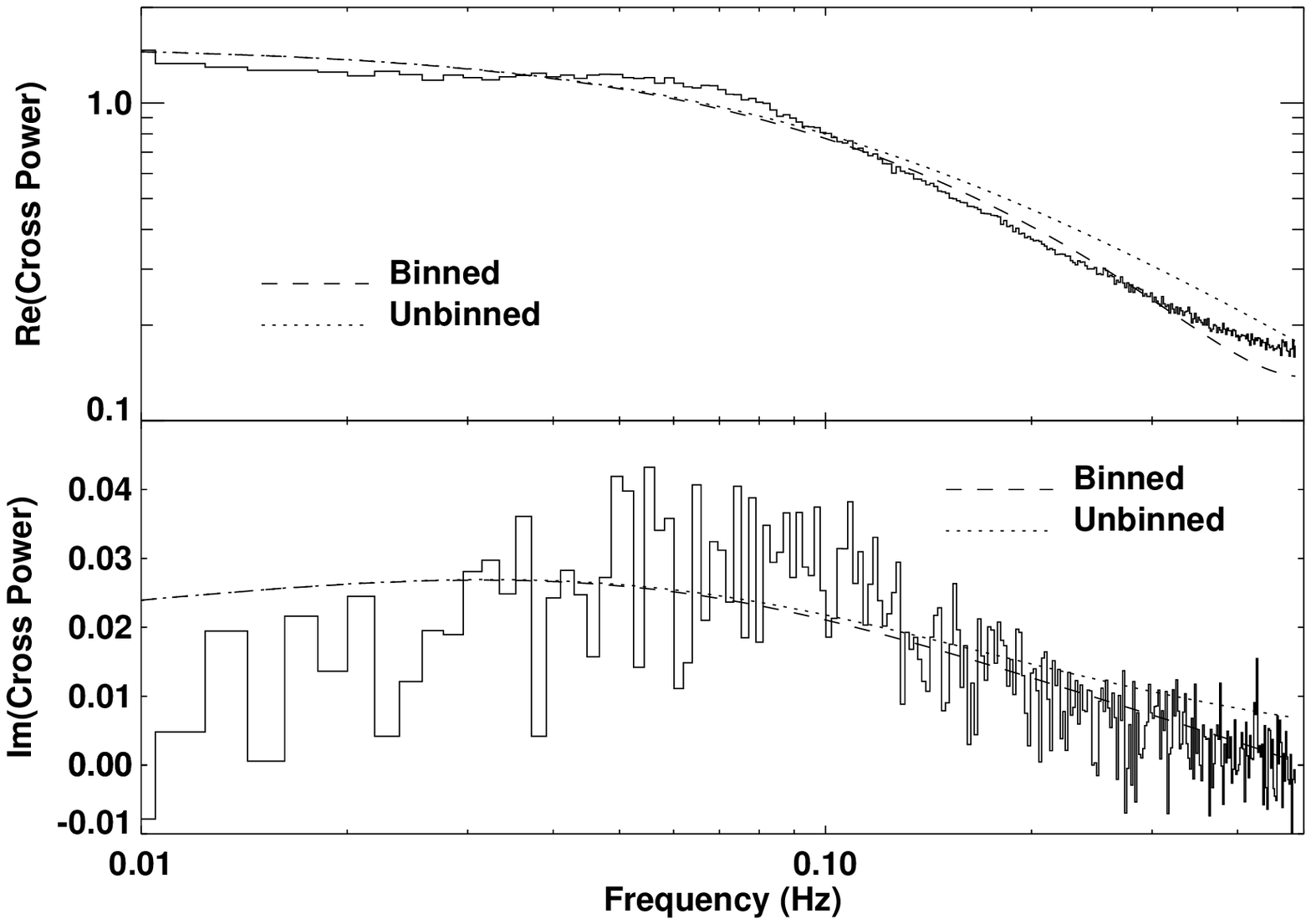,width=6.5in}
\figcaption[crary_fig3.ps]{Real part (upper panel) and imaginary part of
the cross spectrum. The dashed line is a fit of the lag model
to the cross spectrum with the effects of binning included.  The
dotted line is the corresponding quantity without the effects of
binning considered.\label{fig3}}


\newpage
\begin{references}

\reference{Belloni} Belloni, T., \& Hasinger, G. 1990, \aap, 227, L33

\reference{Crary96a} Crary, D.J., et al. 1996, \apjl, 462, L71

\reference{Crary96b} Crary, D.J., et al. 1996, \aaps, 120, 153

\reference{Leahy} Leahy, D. A., Darbro, W., Elsner, R. F.,
Weisskopf, M. C., Sutherland, P.G., Kahn, S., \& Grindlay, J. E.
1983, \apj, 226, 160

\reference{Lewin}Lewin, W.H.G., van Paradijs, J., and van der Klis, M.
1988, \ssr, 46, 273

\reference{Maejima} Maejima, Y., Makishima, M., Matsuoka, Y., Ogawara,Y.,
Oda, M., Tawara, Y., and Doi, K. 1984, \apj, 285, 712

\reference{Miller} Miller, M.C., 1995, \apj, 441, 770

\reference{Miyamoto88} Miyamoto, S., Kitamoto, S., Mitsuda, K., and Dotani, T.
1988, Nature, 336, 450

\reference{Miyamoto89} Miyamoto, S, and Kitamoto, S. 1989, Nature, 342, 773

\reference{Miyamoto92} Miyamoto, S., Kitamoto, S., Iga, S., Negoro, H, and Terada, K.
1992, \apjl, 391, L21

\reference{Miyamoto93} Miyamoto, S., Iga, S., Kitamoto, S. Kamado, Y. 1993,
\apjl, 403, L39

\reference{Nowak} Nowak, M.A., and Vaughan, B.A. 1996, \mnras, 280, 227

\reference{vdK87} van der Klis, M., Hasinger, G., Stella, L. Langmeier, A.,
van Paradijs, J., and Lewin, W.H.G. 1987, \apjl, 319, L13

\reference{vdK89} van der Klis, M. 1989, in Timing Neutron Stars,
ed. H. Ogelman \& E.P.J. van den Heuvel,  (Dordrecht: Kluwer Academic
Publishers), 319

\reference{Paciesas97} Paciesas, W. S. et al., in Proc. Forth Compton
Symposium, Williamsburgs 1997, in press.

\reference{vdK95} van der Klis, M. 1995, in X-ray Binaries, ed. W.H.G. Lewin,
J. van Paradijs, \& E.P.J. van den Heuvel
(London: Cambridge University Press), 252


\end{references}
\end{document}